\documentclass[11pt,twocolumn]{article}

\usepackage{graphicx}      % include this line if your document contains figures
\usepackage{natbib}        % required for bibliography
\usepackage{subfigure}
\usepackage{amssymb}
\usepackage{mathtools}
\usepackage[margin=1.5cm]{geometry}

\title{\bf{Estimating 3D Signals With Kalman Filter} } 
\author{Y.KHMOU ,    S.SAFI , \\
Department of Physics,Polydisciplinary Faculty, Po.Box 592\\
Sultan Moulay Slimane University,Beni Mellal, Morocco.} 

\begin{document}
%\begin{frontmatter}
\maketitle

%\author[Second]{,SAFI Said} 
%\author[Third]{,FAKIR Mohamed}
%\address{Information Processing and Telecommunications Team, Faculty of Sciences and Technics, Sultan Moulay Slimane University, Béni Mellal, Morocco}
%\address[Second]{Department of Physics, Polydisciplinary Faculty, Sultan Moulay Slimane University, Morocco (e-mail: safi.said@gmail.com).}
%\address[Third]{Telecommunication and Information Tehcnology,Faculty of Science and Technology,Beni Mellal, Morocco. (email:fakfad@yahoo.fr).}

\begin{abstract}  
\bf
In this paper, the standard Kalman filter is implemented to denoise the three dimensional signals affected by additive white Gaussian noise (AWGN), we used  fast algorithm based on Laplacian operator to measure the noise variance and a fast median filter to predict the state variable.
The Kalman algorithm is modeled by adjusting its parameters for better performance in both filtering and in reducing the computational load  while conserving the information contained in the signal .              
\end{abstract}

%\begin{keywords}

%\end{frontmatter}
%===============================================================================

\section{\bf Introduction :}
In 1960, RE Kalman published his paper $[1]$ which describes the recursive solution to the problem of linear filtering of discrete data. Since then, the Kalman filter was the subject of extensive research and applications. Stanley Schmidt is recognized as having achieved the first implementation of the filter, later many varieties were used in different fields such as robotic vision and navigation.\\ \\
In image and video processing and denoising techniques, many works have been done lately[2-5] : In $[2]$ the Kalman filter is applied  on noisy static image where the state vector propagates through the image matrix and invoking the orthogonality principle to obtain  the filter parametrs.In $[3]$ the Unscented Kalman filter is used  for incorporating a non-Gaussian prior through importance sampling.while in $[4]$ the filter is enhanced by introducing a factor Gain that controls the quality of filtering for reducing the artifacts.\\ \\
In this paper, we impelemented the standard Kalman filter to denoise three dimensional signals corrupted by additive white Gaussian noise (AWGN), the main contribution in our works is the use of fast method based on Convolution with Laplacaian operator $[6]$ to estimate the noise variance while the state prediction is based on fast median filter $[7]$ .In the next section we explain in details the methodology used in our algorithm.

\section { Proposed Algorithm}
 As one of recent  proposed methods, C.P Mauer [4] uses a frame based method for denoising image sequences obtained by Magnetic Resonance Imaging technique (MRI) or samples taken by Microscope Lapse Time images by introducing a gain factor $G$ in the main Kalman algorithm [1] as described in the following equation :\\
\begin{equation}
X_{estim} = G.X_{pred}+\alpha+  K .(X_{pred}-X_{measur})
\end{equation}\\
\begin{equation*}
\alpha = (1-G).X_{measur}
\end{equation*}

The main advantage in this method is that the wrong guess of the initial variance will not prevent noise estimation but will delay the fitting process, also high values for the filter gain $G$ renders the output less sensitive to momentary fluctuations.\\

In the proposed algorithm we consider  the state variable the snapshot (instantaneous) image  variable $ X(x,y,t_k)$ that propagates through time t with $ t_{k}=[t_0,....t_f]$ .\\
In the other hand , we consider the  two stochastic processes : 
\begin{equation}
X_{n+1}= A. X_{n}+ W_{n} 
\end{equation}

\begin{equation}
Y_n =H .  X_{n}+ N_{n}
\end{equation}

With {$N$,$W$} represent process noise  and error measurment covariances respectively  with parameters   :
\begin{equation*}
W_{n}\sim N(0,Q_{n})
\end{equation*}
\begin{equation*}
N_{n} \sim N(0,R_{n})
\end{equation*}

Modelling the two processes described by the equations $2$,$3$ to fit the system of three dimensional noisy video signals yields to the following new representation :

\begin{equation}
X_{n+1} = X_{n}+ W_{n}
\end{equation}

\begin{equation}
Y_{n}=X_{n}+ N_{n}
\end{equation}

$Q_n$ :the variance process which represents the changing between two successive frames.\\
$R_n$ :the noise variance  such as for every pixel $x_{i,j}$ its correspondant couple $(q_{i,j},r_{i,j})$ 

$ (X_n,Y_n)$ forms a hidden Markov process that for estmating the state based on the measurments $Y_n$ requires the computation of the conditional probability $P(X_n|Y_n)$

we consider the the variance process and the noise variance stationary :
\begin{equation}
{{\partial Q} \over {\partial t}} =0 , { {\partial W} \over {\partial t }} = 0
\end{equation}

As we are conserned with fast methods, we made a hypothesis that the two variance matrices  are related to each other linearly :
\begin{equation}
Q=\beta . R
\end{equation}

After many experiences for filtering noisy sequence based on the PSNR value and on changing the value $\beta$ we found that the two matrices must be equals $ Q=R$ to get the higher value of PSNR .
and we define theirs static values using fast method for Gaussian noise [6], we give,  briefly , in the following, the algorithm  note that we took only one part of the algorithm  of [6] in our work.

\subsection{Fast Gaussian noise estimation :}
Given a two dimensional signal $I(n,m)$ the standard deviation of the AWGN is obtained by two main operations which are convolution and averaging :

\begin{equation}
\sigma _n = \sqrt{\pi \over 2}  {1 \over {6(n-2)(m-2)}} {\sum_{x,y \in \Omega}^{} |I(x,y)*N}|
\end{equation}\\
with  $N$ the matrix defined as the following:

\begin{equation}
 N=  \begin{pmatrix}      1 &  -2 &  1 \\ 
                               2 &  4 &  -2 \\
                               1 &  -2 &  1 \end{pmatrix} 
\end{equation}

To evaluate the efficiency of this method, we tested the algorithm on the image "liftingbody.jpg" of size $512*512$to estimate the noise added with zero mean and  20 different values of  standard deviation  $\sigma=[0.0316,...,0.3162]$, the following figure illustrates the result :

\begin{figure}[ht]
%\begin{center}
\includegraphics[width=8.00cm]{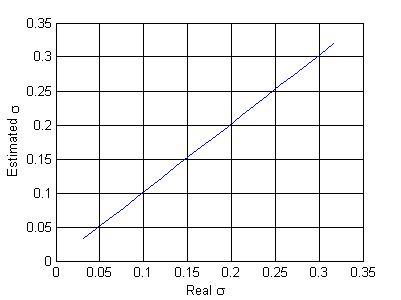}    % The printed column width is 8.4 cm.
\caption{ Estimated Standard deviation } 
\label{fig:figure1}
%\end{center}
\end{figure}
The Root Mean Square Error for this experiment was :
$rmse=0.0021$

Therefore given a manifold $X(x,y,T=[t_1,...,t_f])$ , for every dynamic element $p(i,j,:)$ with {$ 1\leq i \leq x, $ $ 1 \leq j \leq y$} we estimate the noise variance trough the whole signal such as :
\begin{equation}
q_{i,j}=r_{i,j}={\sigma (p_{i,j,1:T})}^2
\end{equation}

The state prediction at time $t+1$ is done using fast median filter $[7]$ with window size $W=3$ such that every frame is transform into one dimensional column wise vector using the mapping operator $vec{.}$,after applying the filter, the result is reshaped into its initial 2D form.The 
initial conditions are computed using the following steps : \\
1.Computing the first state prediction : \\
\begin{equation}
X_{prd} = E[X(x,y,t=1)] \triangleq FastMed[X(x,y,t=1)] 
\end{equation}
2. Estimating  the noise variance  and the process variance  stationary matrices $Q$ and $R$. \\
3. Computing the fist variance error $P(x,y,t=1)$: \\
\begin{equation}
P(x,y,t=1)=Cov[ X_{prd}(x,y,t=1)-X(x,y,t=1)] 
\end{equation}
\begin{equation}
P(x,y,t=1)=E[Err.Err^{T}]
\end{equation}\\
Note that if the signal is not square in spatial dimensions , an adjustment should give the matrix $P$ the same dimensions as every matrix in our system by padding elements or cutting them. \\
4. Compute the Kalman Gain : \\
\begin{equation}
K={{P_{prd}} \over {P_{prd}+Q}}
\end{equation} \\
Note that matrix product used in the whole paper is the Hadamard(elementwise) product.\\
We explain the state processing through time : First, we predict the state at time $t+1$,second, we take the measure at time $t-1$,next,we estimate the state variable ( 2D signal, image, snapshot,..) at time $t$,so the mean operations in our algorithm are presented by the three following equations  : \\

\begin{equation}
X_{prd}=FastMedian[X_{x,y,t+1}] 
\end{equation}

\begin{equation}
 X_{measu}= X_{x,y,t-1}
\end{equation}

\begin{equation}
X_{x,y,t}^{est}=X_{prd}+ K.(X_{measu}-X_{prd})
\end{equation}\\

The rest of the algorithm  for computing the Kalman Gain, the predicted and estimated Variance error, is the same as described in details in $[1]$ .
\subsection{Recapitulation}
We resume in this subsection the main operations in our algorithm :  Given a signal $X(x,y,T)$  affected by AWGN we estimate the true signal using the Kalman algorithm by :\\
1. Computing the inital state variables $P_{prd}$, $K$, $X_{prd}$  .\\
2. Estimating the stationary matrix $Q(x,y)$ and $R(x,y)$ .\\
3. Looping through the third dimension using the protocol :  Prediction at $(t+1)$ , measuring at $(t-1)$ and  estimating at time $(t)$ using the equation $16$.\\

\subsection {Evaluation Criteria }
We used three metrics to evalate the quality of the filtering which are, Mean Square Error(MSE), Peak-Signal-to-Noise-Ratio and the Autocorrelation function(ACF) .The mean square error is given by the following equation :

\begin{equation}
MSE(X,Y)=\frac{1}{NM}\sum_{i=1,j=1}^{i=M,j=N}{({x_{i,j}-y_{i,j}})^2}
\end{equation}

The Peak-Signal-to-Noise-Ratio (PSNR) is defined,in $dB$, by the  following equation :
\begin{equation}
\mathit{PSNR(X,Y)}=10.\log _{10} \left (\frac D{\mathit{MSE(X,Y)}^2} \right )
\end{equation}
Where the metric $D$ is defined by : \\
\begin{equation*}
D=Max[max(X[:]),max(Y[:])]
\end{equation*}

For a signal $X$ sampled $p$ times  $[1,p]$ (containing $p$ slices) we compute the PSNR of  each snapshot $1 \leq i \leq p $ next we take the average result :
\begin{equation*}
\hat{PSNR}={\frac 1{P}} \sum_{n=1}^{p}PSNR_{n}
\end{equation*}
The same process is done to the MSE.
We also measured statistically the removed noise by the autocorrelation function based on Wiener-Khinchin-Einstein theoerem, the ACF of the removed noise must have the statistical carateristics as that of the AWGN : \\
%\begin{thm}
% Wiener-Khinchin-Einstein Theorem : 
%The power spectral density of wide-sense-stationary-random process is the Fourier %transform of the corresponding auto-correlation function : \\
%
%\end{thm}
The autocorrelation function $R_{x}(u,v)$ is related to the power spectral density (PSD) via the following equation :
\begin{equation}
S_{x}(a,b)= \int_{-\infty}^{+\infty}\int_{-\infty}^{+\infty} R_{x}(a,b)\exp ^{(-2\pi i.(au+bv))} du.dv %= \mathit {F}[R_{x}(a,b)]
\end{equation}
On the other hand , the power spectral density is the square modulus of the Fourier transform of the 2D signal X :
\begin{equation}
S_x= \|F[X]  \| ^2 = F[X].\overline{F[X]}
\end{equation}\\
Then, the fast equation of the autocorrelation function :
\begin{equation}
R(X)= F^{-1}[F[X].\overline{F[X]}]
\end{equation}

A special case is when the signal is (AWGN), the autocorrelation becomes  :
\begin{equation}
r(n,m)= \sigma^2 .\delta(n,m) 
\end{equation}

$\sigma$ : constant power spectral density.
$\delta$ : 2D Dirac implusion .
An illustrative result is presented in figure 8 in this next section.

\section{\bf SIMULATION Results :}

The simulation of the proposed algorithm was done using the software $MATLAB^{TM}$ R2007a, different signals were tested including  MRI ( Magnetic Resonance Imaging),DICOM (Digital Imaging and Communications in Medicine) sequences, gray scale videos acquired via Universal Serial Bus (USB) WebCam  and standard grayscale videos .\\
We represent in the figure 2, the result of filtering a time lapse images obtained on a Zeiss LSM510 cofocal microscope[4] :

\begin{figure}[ht]
%\begin{center}
\includegraphics[width=8.00cm]{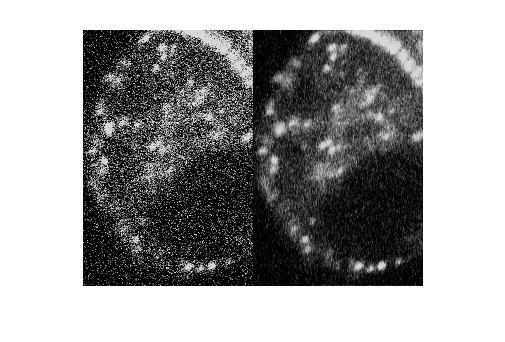}    % The printed column width is 8.4 cm.
\caption{ left panal : $38^{T}$ Frame of  raw data with  estimated noise deviation $\sigma=0.2$ , right panel : filtered frame  } 
\label{fig:figure2}
%\end{center}
\end{figure}
After simulating noise of variance $\sigma^{2}= [0.001...0.1]$ applied to the scene 'gmissa.avi' with dimensions $X(288,360,150)$ of normalized pixel values $[0,1]$ We represent  some results with $\sigma=0.06$ :
The average  MSE between the noisy and the filtered signals was : $MSE = 0.0034$ while the same metric between the filtered and original signals was :$MSE=7.7665e-004 $
\begin{figure}[ht]
%\begin{center}
\includegraphics[width=8.00cm]{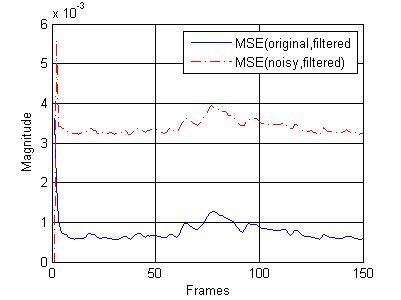}    % The printed column width is 8.4 cm.
\caption{ Mean Square Error (noisy,orignal,filtered)  } 
\label{fig:figure3}
%\end{center}
\end{figure}

Figure 4 in left represents the noisy $26^{th}$ frame taken from 'gmissa.avi" scene which has $PSNR=25.22 dB$, while the corresponding right frame represents the result of filtering which has $PSNR=31dB$, our algorithm is capable of reducing the noise with the difference of $5.78 dB$ which is considered  acceptable .

\begin{figure}[ht]
\mbox{\subfigure{\includegraphics[width=5cm]{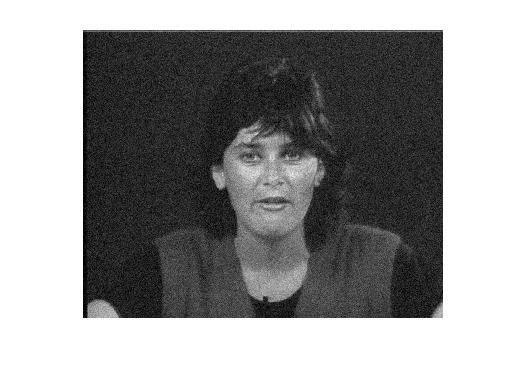}}  \hspace{-1cm}
      \subfigure {\includegraphics[width=5cm]{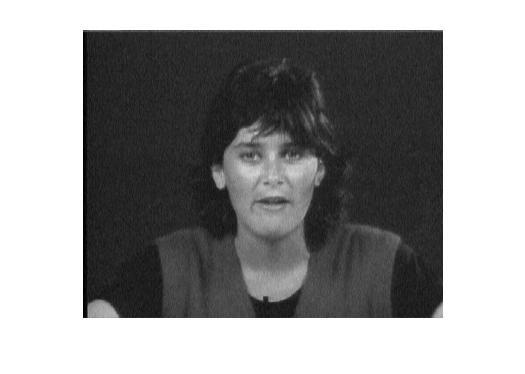}}}
      \caption{gmissa scene}
      \label{fig4}
\end{figure}

Figures 5 and 6 represent the average dynamics of the Kalman filter algorithm which are the mean error variance and the mean Kalman gain  such that for $t=1:150$ :

\begin{equation}
K_{t}={\frac{1}{288*360}} \sum_{i=1,j=1}^{288,360} K_{i,j,t}
\end{equation}

\begin{equation}
P_{t}={\frac{1}{288*360}} \sum_{i=1,j=1}^{288,360} P_{i,j,t}
\end{equation}

We note that the Kalman gain  convergs to a final value starting from  the 7th frame and conserves the same value for the remaining 147 frames . 

\begin{figure}
%\begin{center}
\includegraphics[width=8.00cm]{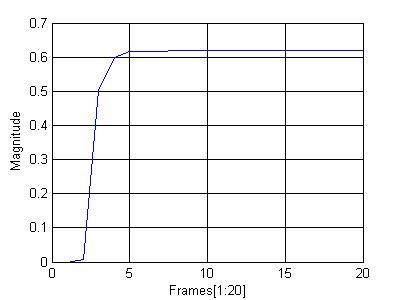}    % The printed column width is 8.4 cm.
\caption{Mean  Kalman gain over 20 frames} 
\label{fig:figure5}
%\end{center}
\end{figure}

\begin{figure}
%\begin{center}
\includegraphics[width=8.00cm]{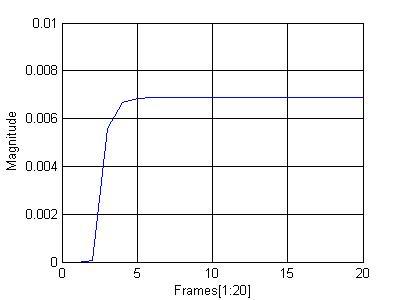}    % The printed column width is 8.4 cm.
\caption{Mean Error variance matrix over 20 frames, convergence attained in the 5th frame  } 
\label{fig:figure6}
%\end{center}
\end{figure}

The figure 6 represents the  dynamic PSNR evaluated over all frames to verifiy the degree of stability of the algorithm :
\begin{figure}[ht]
%\begin{center}
\includegraphics[width=8.00cm]{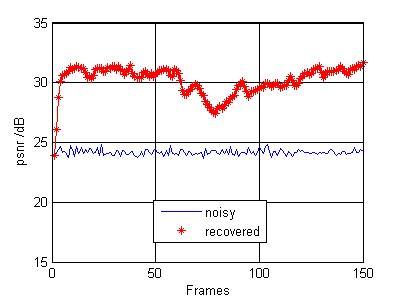}    % The printed column width is 8.4 cm.
\caption{PSNR of noisy and filtered signals with noise std $\sigma =0.05$ , $E[PSNR_{noisy}]= 24.20 dB $ and $E[PSNR_{filtered}]=31.19 dB$ .} 
\label{fig:figure7}
%\end{center}
\end{figure}

In the figure 8, we show the result of the two dimensional Autocorrelation Function applied, using the equation $22$, to the residual from the $48^{th}$ frame, the structure is the ACF is conformal to the equation $23$ .

\begin{figure}[ht]
%\begin{center}
\includegraphics[width=8.00cm]{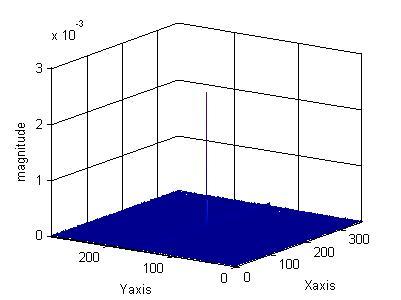}    % The printed column width is 8.4 cm.
\caption{Two dimensional autocorrelation function of residual n° 48 in gmissa scene,$Peak=0.0033$} 
\label{fig:figure8}
%\end{center}
\end{figure}
Results of the scene "gflower" affacted by $\sigma=0.1095$ :
\begin{figure}[ht]
\mbox{\subfigure{\includegraphics[width=5cm]{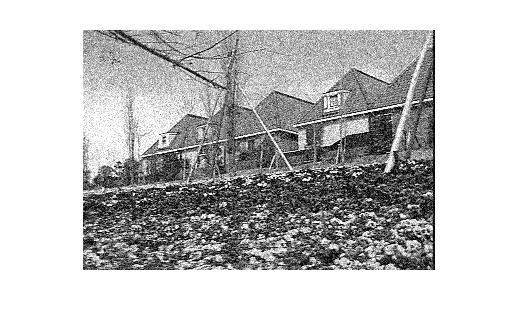}}  \hspace{-1cm}
      \subfigure {\includegraphics[width=5cm]{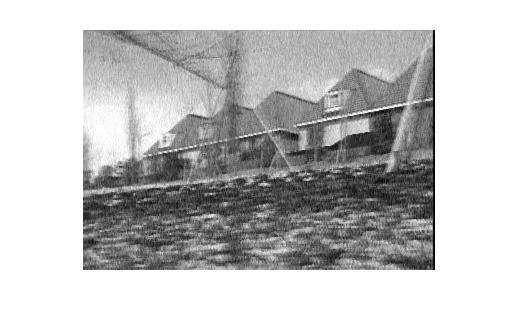}}}
      \caption{Left :noisy Frame 100 with $PSNR=17.77 dB$,right :filtered 100  
               frame with$PSNR=22.106dB$}
      \label{fig9}
\end{figure}
The following table illustrates some of results obtained using different signals :
\begin{table}[htbp]
\caption{PSNR for differents video signals with varying noise $\sigma$.}
\begin{tabular}{|p{0.5cm}|p{1.75cm}|p{1.75cm}|p{1.75cm}|p{1.75cm}|}
\hline
\multicolumn{1}{|l|}{$\sigma$}  & costguard & gsalesman & gstennis & USB Camera. \\ \hline
0.03 & 25.66/30.59 & 28.25/30.14 & 22.00/30.86 & 28.54/30.75 \\ \hline
0.15 & 19,19/22,88 & 17.98/22.61 & 20.03/21.24 & 20.18/24.59 \\ \hline
0.21 & 17.37/21.53 & 16.54/21.03 & 18.31/20.67 & 16.92/18.66 \\ \hline
0.25 & 16.49/20.68 & 15.71/20.05 & 17.39/20.23 & 16.31/18.32 \\ \hline
0.31 & 15.67/19.73 & 15.34/19.12 & 16.56/19.64 & 17.12/20.99 \\ \hline
\end{tabular}
\label{table1}
\end{table} \\ \\ \\

\section{Conclusion}
In this paper, a modified version of Standard Kalman Filter was applied on three dimensional signals corrupted by Additive White Gaussian Noise (AWGN), we used fast method based on Laplacian operator to estimate noise standard deviation, and fast median filter for state prediction such that configured the Kalman algorithm to be frame based method. Exprimental results showed an improvement of approximatly $5.5 dB$ which makes this approach considerable in comparaison with sophisticated filtering techniques .

As future perspective we will try generalize the algorithm for 4D Signals especially the  RGB video signals, and we will try to increase the PSNR with fast method to estimate the variance process between each two consecutive frames in the signal .

%\appendix
%\section{A summary of Latin grammar}    % Each appendix must have a short title.
%\section{Some Latin vocabulary}          
\end{document}